\documentstyle[12pt,aaspp4]{article}
\tighten

\lefthead{Band}
\righthead{Burst Repetition}

\begin{document}

\title{SKY COVERAGE AND BURST REPETITION}
\author{David L. Band}
\affil{CASS 0111, University of California, San Diego, La Jolla, CA  92093}
\centerline{\it Received 1995 September 14; accepted 1996 January 5}
\centerline{To appear in the June 20, 1996, issue of}
\centerline{\it The Astrophysical Journal}
\begin{abstract}
To investigate the repeater content of gamma ray burst samples I develop two
models where sources burst at a constant average rate.  I find that the sky
coverage affects the number of repeaters in a sample predominantly through the
detector livetime, and that the number of bursts in the sample is the primary
parameter.  Thus the repeater content of burst samples should be compared
within the context of a repetition model; a direct comparison between two
samples is possible only if the samples have similar sizes.  The observed
repeater fraction may not be the actual fraction if the sources burst on
average less than once during the detector livetime. Sources which burst
repeatedly during active phases separated by more than the observation period
must be treated separately. 
\end{abstract}
\section{INTRODUCTION}
Whether the observations from the Burst and Transient Source Experiment (BATSE)
on the {\it Compton Gamma Ray Observatory} ({\it CGRO}) show that classical
gamma ray bursts repeat is of great current interest because most of the models
for bursts at cosmological distances destroy the burst source in producing the
burst.  The reports of repeaters in BATSE's 1B (i.e., first) catalogue of 260
bursts (Fishman et al. 1994) were based on spatial (Quashnock \& Lamb 1993) and
spatial-temporal (Wang \& Lingenfelter 1993, 1995) clustering. The spatial
clustering is not evident in the 2B (i.e., second) catalogue (which includes
the bursts of the 1B catalogue---Meegan et al. 1994) of a total of 585 bursts
(Meegan et al. 1995a), although whether a weak spatial-temporal clustering
signal is present is disputed (Brainerd et al. 1995 do not find the signal that
Wang \& Lingenfelter 1995 and Petrosian \& Efron 1995 find).  All repeater
signals are absent using the revised burst positions in the 3B catalogue
(Meegan et al. 1995b), even for the bursts which were included in the 1B
catalogue. While the BATSE observations do not show convincingly that burst
sources repeat, they do not rule out such repetitions.  Because of the large
uncertainties in the positions of bursts localized by BATSE (less than
2$\arcdeg$ systematic and from less than 1$\arcdeg$ to more than 15$\arcdeg$
statistical, depending on the burst intensity---Meegan et al. 1995b), studies
of the repeater content of the BATSE catalogs generally do not identify
individual repeating sources but instead determine whether the spatial 
clustering is greater than expected for an isotropic distribution of events, or
alternatively what fraction of the observed events could be from repeaters. 
These statistical methods can only address what is actually in the data.
Explicit or implicit in deriving the physical implications of these analyses
are assumed models of source repetition.  Searches for spatial-temporal
clustering assume that burst sources repeat during active phases while the
searches for spatial clustering are sensitive to any repetition pattern.  Here
I investigate the models behind previous studies, and develop two simple models
of burst repetition.  Note that I explore the connection between the observed
and actual source repetitions, and not the accuracy in determining a burst
sample's observed repeater content. 

The reduced sky coverage during the second year of BATSE's operation has been
invoked to explain the weak or absent repeater signal during BATSE's second
year under the assumption that burst sources repeat.  The bursts in the 2B
catalogue but not in the 1B (usually referred to as 2B-1B) were accumulated
during this second year. I define sky coverage $f_s$ as the fraction of the sky
BATSE views on average; the sky coverage is less than 1 as a result of Earth
blockage, SAA passages, telemetry gaps, etc.  Indeed, during BATSE's first year
$f_s\sim1/3$ and during the second $f_s\sim1/4$.  Some of the repeater studies
model the repetitions among the observed bursts without regard for the actual
repeater pattern (Strohmayer, Fenimore \& Miralles 1994; Meegan et al. 1995a;
Brainerd et al. 1995), while others model the repetitions in the underlying
burst population, whether observed or not. Here I argue that deriving a
physically meaningful statement regarding the presence of repetitions depends
on a model of such repetitions.  In particular, the comparison of different
burst ensembles (e.g., the 1B vs. 2B-1B catalogues) requires an assumed
repetition model. 

To make this analysis more concrete, I first develop in detail two simple
models: a ``stochastic'' model where the probability of a source bursting per
unit time is constant (\S 2) and a model where the source bursts at a constant
rate (\S 3).  Based on these models, I comment on features of more general
repeater models (\S 4).  Finally I discuss the use of repetition models in
studies of the repeater content of the BATSE bursts (\S 5).  Great care must be
taken in defining quantities so that it is clear whether a quantity includes
all bursts during an observation period or only those which are observed, and
whether a repeating source is one which eventually will repeat or one actually
observed to repeat. 
\section{STOCHASTIC BURSTS} 
Assume that each source bursts stochastically at a fixed rate (this is a
modeling assumption): the probability of an observable burst within any given
time interval is constant and independent of the intensity of, or time since,
the previous burst. Since detector thresholds differ, the rate is
detector-dependent.  Similarly, subsets of a burst sample with different
thresholds have different effective rates. A source that bursts only once
(i.e., does not repeat) can be treated as having a rate of once per Hubble
time.  Thus the number of bursts observed from the $i$th source which bursts at
a rate $r_i$ over a time $\Delta T$ by a detector with a sky coverage $f_s$
will have a Poisson distribution with mean (and variance) $n_i=r_i f_s \Delta
T$.  In this model it does not matter whether the detector's livetime is
contiguous ($f_s=1$) or highly chopped up ($f_s \ll 1$); only the total
livetime $\tau=f_s \Delta T$ is relevant. 

For a fixed number of sources $N_S$, the number of bursts observed during a
given livetime $\tau$ is 
\begin{equation}
\langle N_B \rangle = \sum_{i=1}^{N_S} \sum_{n=1}^\infty 
   n {{(r_i \tau)^n}\over{n!}} \exp[-r_i \tau] 
= \sum_{i=1}^{N_S} r_i \tau
\end{equation}
while the number of bursts from sources which are observed to repeat (i.e.,
from which two or more events are observed) is 
\begin{equation}
\langle N_{B,r} \rangle = \sum_{i=1}^{N_S} \sum_{n=2}^\infty 
   n {{(r_i \tau)^n}\over{n!}} \exp[-r_i \tau] 
= \sum_{i=1}^{N_S} r_i \tau (1-\exp[-r_i \tau])
\end{equation}
and the fraction of all events which are observed to originate from repeating
sources is
\begin{equation}
f_r = {{\langle N_{B,r} \rangle}\over {\langle N_B \rangle}}   = 
   {{\sum_{i=1}^{N_S} r_i \tau (1-\exp[-r_i \tau])}\over
   {\sum_{i=1}^{N_S} r_i \tau}} \quad .
\end{equation}
Thus $N_B$ is proportional (on average) to $\tau$.  For those sources with
$1/r_i \le \tau$ few repeaters will be observed ($n_i\le 1$ and the Poisson
probability of 2 or more events will be small).  Both the average number of
repetitions within a burst sample and the size of the burst sample are
functions of $\tau$, and thus a sample's observed repeater content is a
function of its size.  Note that as $\tau$ and the sample size increase, more
sources will be observed to repeat and the burst sample will include more
repetitions.  Of course the ability to detect repetitions in the burst sample
may decrease as the sample size increases.  For example, the angular
correlation function is inversely proportional to the number of bursts $N_B$: 
$w(\theta) \propto [(N_B-1)\sigma_*^2]^{-1}$, where $\sigma_*$ is the
uncertainty in the burst positions (Meegan et al. 1995a). 

Note that the product $r_i \tau$ recurs in all expressions for the number of
observed repetitions, bursts observed, etc.  In reality, the sky coverage $f_s$
and thus the livetime $\tau$ are not constant over the sky. The local value of
$\tau_i$ (i.e., the value of $\tau$ at the location of the $i$th source) is the
relevant parameter, but the above argument can be extended from a distribution
of $r_i$ to a distribution of $r_i \tau_i$.  Here I will neglect the variations
of the sky coverage over the sky.  Hakkila et al. (1995) use a sky coverage
which varies with declination to constrain the repeater fraction from the
observed isotropy assuming a stochastic model.

As an example, assume that there are $N_S$ repeating sources with the same 
burst rate $r$ as well as a class of nonrepeating sources.  The nonrepeating
sources can be treated as contributing bursts at a constant rate $R$.  Then
\begin{eqnarray}
\langle N_B \rangle &=& (R+N_S r)\tau \quad , \\
\langle N_{B,r} \rangle &=& N_S r\tau (1-\exp[-r\tau]) \quad , \\
\hbox{and} \qquad f_r &=& F_r (1-\exp[-r\tau]) \qquad \hbox{where} \qquad 
F_r={1\over {1+(R/N_S r)}} 
\end{eqnarray}
is the asymptotic value of $f_r$ for
large $\tau$ when all the repeating sources will have been observed to burst at
least twice.  This asymptotic $F_r$ is the physically meaningful repeater 
fraction which analysis of the burst samples should strive to determine.  The 
number of sources observed to repeat is 
\begin{eqnarray}
\langle N_{S,obs} \rangle =& 
   N_S \left(1-\left(1+r\tau \right) \exp[-r\tau] \right) &\\
   =& N_S (r\tau)^2 /2 & \hbox{for}\qquad \tau \rightarrow 0 \nonumber\\
   =& N_S & \hbox{for}\qquad \tau \rightarrow \infty \quad .\nonumber
\end{eqnarray}
At large livetimes $\tau$ every repeating source is indeed observed to repeat. 

The average number of events observed from sources observed to repeat (i.e.,
excluding bursts from repeating sources which are observed to burst only once)
is
\begin{eqnarray}
\langle n_{obs} \rangle =& {{\sum_{n=2}^\infty n 
   {{(r\tau)^n}\over{n!}}\exp[-r\tau]}\over{\sum_{n=2}^\infty 
   {{(r\tau)^n}\over{n!}}\exp[-r\tau]}} =&
{{r\tau (1-\exp[-r\tau] )}
\over {\left(1-\left(1+r\tau \right) \exp[-r\tau] \right)}} \\
   =& 2  &\hbox{for}\qquad \tau \rightarrow 0 \nonumber\\
   =& r\tau  &\hbox{for}\qquad \tau \rightarrow \infty \quad .\nonumber
\end{eqnarray}
For small livetimes only two events will be seen from the few sources observed
to burst more than once, but eventually all repeating sources will be observed
to burst many times. 

Strohmayer et al. (1994) used a different repeater fraction which excludes the
first burst of each series of two or more bursts from a repeating source.
The number of observed bursts which are repetitions of a previously observed
burst is $\langle N_{B,r} \rangle - \langle N_{S,obs} \rangle$.  The resulting 
fraction is 
\begin{equation}
f_r^\prime = {{\langle N_{B,r} \rangle - \langle N_{S,obs} \rangle}\over
   {\langle N_B \rangle}} = 
   F_r \left( 1-\left(1-\exp[-r\tau]\right)/r\tau \right) \quad .
\end{equation}

Figure~1 shows these various quantities as a function of the expected number of
observed bursts per source ($r \tau$) for a model where all sources repeat at a
constant average rate $r$ (i.e., there are no nonrepeating sources and
$F_r=1$).  The total number of bursts in the ensemble is linearly proportional
to $r\tau$. 

Note that in this model the repeater content of a burst sample---the number of
sources observed to repeat---is a function the sky coverage $f_s$ only through
the dependence of the livetime $\tau$ on $f_s$.  The number of bursts observed
is linearly proportional to the livetime and can be used as a surrogate for the
livetime.  I therefore expect the repeater character of two samples with
comparable numbers for bursts to be the same.  Conversely, I expect different
observed repeater characteristics for samples with different sizes. 
\section{BURSTS AT A CONSTANT RATE}
Here I assume that the $i$th source bursts a fixed number of times $n_i$ during
a given observation period $\Delta T$.  The sky coverage $f_s$ is the
probability that any one of these bursts will be observed.  The number of
bursts observed from any source is characterized by the binomial distribution:
the probability of observing $n\le n_i$ bursts is $P(n) =
{{n_i!}\over{n!(n_i-n)!}} f_s^n (1-f_s)^{n_i-n}$.  Thus the quantities defined
in \S 2 are 
\begin{eqnarray}
\langle N_B \rangle &=& \sum^{N_S}_{i=1} \sum_{n=1}^{n_i} n 
   {{n_i!}\over{n!(n_i-n)!}} f_s^n (1-f_s)^{n_i-n} 
   = \sum^{N_S}_{i=1} n_i f_s \\
\langle N_{B,r} \rangle &=& \sum^{N_S}_{i=1} \sum_{n=2}^{n_i} n 
   {{n_i!}\over{n!(n_i-n)!}} f_s^n (1-f_s)^{n_i-n} 
   = \sum^{N_S}_{i=1} n_i f_s (1-(1-f_s)^{n_i-1}) \\
f_r &=& {{\sum^{N_S}_{i=1} n_i f_s (1-(1-f_s)^{n_i-1}) }\over
   {\sum^{N_S}_{i=1} n_i f_s }} \\
\langle N_{S,obs} \rangle &=& \sum^{N_S}_{i=1} \sum_{n=2}^{n_i} 
   {{n_i!}\over{n!(n_i-n)!}} f_s^n (1-f_s)^{n_i-n}  \\
   &=& \sum^{N_S}_{i=1} \left( 1- (1-f_s)^{n_i} - 
   n_i f_s (1-f_s)^{n_i-1} \right) \nonumber \\
\langle n_{obs} \rangle &=& {{\sum^{N_S}_{i=1} \sum_{n=2}^{n_i} n 
   {{n_i!}\over{n!(n_i-n)!}} f_s^n (1-f_s)^{n_i-n} }\over
   {\sum^{N_S}_{i=1} \sum_{n=2}^{n_i} 
   {{n_i!}\over{n!(n_i-n)!}} f_s^n (1-f_s)^{n_i-n} }} 
   = {{\langle N_{B,r}\rangle}\over{\langle N_{S,obs} \rangle}} \quad .
\end{eqnarray}
As in \S 2, $\langle N_B \rangle$ is the number of observed bursts, while
$\langle N_{B,r} \rangle$ is the number of observed bursts which originate on
sources with more than one observed burst. These $\langle N_{B,r} \rangle$ 
bursts constitute a fraction $f_r$ of the observed bursts $\langle N_B \rangle$.
The fraction of all bursts, observed or not, produced by repeating sources
is $F_r$; asymptotically $f_r$ tends to $F_r$.  The number of sources observed
to repeat is $\langle N_{S,obs} \rangle$ and these sources are observed to 
burst $\langle n_{obs} \rangle$ times.

If there are nonrepeating sources which provide bursts at a rate $R$ and $N_S$
repeating sources which each burst $n_i=n_B$ times during the observation
period (the model used by Quashnock [1995a,b] and Graziani [1995]), then 
\begin{eqnarray}
\langle N_B \rangle &=& f_s(R\Delta T + N_S n_B) \quad , \\
\langle N_{B,r} \rangle &=& N_S f_s n_B (1-(1-f_s)^{n_B-1}) \quad , \\
f_r &=& F_r \left( 1-\left( 1-f_s \right)^{n_B-1} \right) \qquad \hbox{where}
   \qquad F_r = {1\over{1+(R\Delta T / N_Sn_B)}} \quad ,\\
\langle N_{S,obs} \rangle &=& N_S \left( 1- (1-f_s)^{n_B} - 
   f_s n_B (1-f_s)^{n_B-1} \right) \quad , \\
\langle n_{obs} \rangle &=& 
   {{f_s n_B (1-(1-f_s)^{n_B-1})}\over
   {1- (1-f_s)^{n_B} - f_s n_B (1-f_s)^{n_B-1} }} \quad .
\end{eqnarray}
Note that the effective rate is $r=n_B/\Delta T$, and the number of bursts
observed from repeating sources is $f_s n_B N_S = f_s N_S r\Delta T = N_S r
\tau$ where $\tau=f_s\Delta T=f_s n_B/r$ is the livetime.  Although there is
formally a dependence on $f_s$ beyond its effect on the livetime, the
quantities $f_r$, $\langle N_{S,obs} \rangle$ and $\langle n_{obs} \rangle$ are
very nearly the same functions of $f_s n_B$ (which is proportional to the
livetime).  Figure~2 shows these quantities as a function of $f_s n_B$ for
$f_s=1/4$ and $f_s=1/3$. The x-axes of Figures 1 and~2 are the same (the 
number
of bursts observed from each repeating source), and the curves for these three
models are almost identical above one burst observed per repeating source. 
\section{MORE GENERAL MODELS} 
The large range of possible repetition models makes a systematic analysis
impossible, but general comments can be made about classes of models. Barring
correlations between the sky coverage and the repetition pattern, the repeater
content of a burst ensemble is primarily a function of the size of the ensemble
for models where the source bursts at an average rate. We showed in detail for
two models with an average burst rate that the dependence on the sky coverage
is felt through its effect on the livetime.  Complications arise when the
repetition pattern has a time scale commensurate with a characteristic time of
the sky coverage.  For example, if a repetition tends to occur $\sim$45 min
(i.e., half an orbit) after an initial burst, the Earth may occult one of the
bursts for a detector in low Earth orbit. 

One might have expected a stronger dependence on the sky coverage $f_s$ which
is the probability of observing a given burst.  Although the probability of
seeing both members of a pair of bursts from a repeating source is $f_s^2$, the
quantity of interest is the conditional probability of observing repetitions
from the source of a previously observed burst. Assuming a repeating source
bursts twice during the observation period $\Delta T$, the conditional
probability of detecting a repetition of an observed burst is $f_s$.  The
sample size accumulated is proportional to $\tau=f_s \Delta T$.  Decreasing
$f_s$ does reduce the probability of observing a repeater, but it also
decreases the number of bursts in the sample.  To maintain the sample size,
$\Delta T$ must be increased as $f_s$ is decreased.  But a key assumption was
that there were only two bursts during $\Delta T$; if $\Delta T$ increases then
the effective burst rate $r=2/\Delta T$ decreases. 

Therefore for $f_s$ to be the probability of observing the repetition of a
given burst in a burst sample of a specified size, the effective burst rate
must be proportional to the inverse of $\Delta T$. Clearly this is not the case
for any repetition model where the rate is constant. However, if the source
goes through active bursting phases shorter than $\Delta T$, and the average
separation between the active phases is longer then $\Delta T$, then $r\propto
\Delta T^{-1}$.  This is the class of models which Wang \& Lingenfelter (1993,
1995) consider.  For example, if a source bursts twice within five days, and
then does not burst again for a decade, then $r=2/\Delta T$ for $\Delta T\sim$1
yr.  However, as $\Delta T$ increases we expect an increase in the number of 
repeating sources which become active during the observation period.

Similarly studies of the sky coverage-dependence of observing repetitions from
sources with a fixed number of bursts within the observation period $\Delta T$
implicitly assume the burst rate is inversely proportional to $\Delta T$.  This
assumption excludes fixed rate models, but permits active phase models. 
\section{DISCUSSION}
Above I considered repetitions in the underlying source population, and
calculated their observed consequences.  This approach provides a basis for
comparing different burst samples for consistency in the apparent repeater
content, and thus is necessary for resolving whether burst sources repeat.  In
addition, this formulation permits observational results (e.g., upper limits on
the observed repeater fraction $f_r$) to be translated into physical
constraints on bursts sources (e.g., limits on the rate at which sources burst
and the asymptotic repeater fraction $F_r$).  As stated in the Introduction,
because of the large uncertain in burst positions, most studies do not
attribute individual bursts to specific repeating sources, but instead place
statistical constraints on the repeater fraction.  I found that the observed
repeater content for models with a constant average burst rate is predominantly
a function of the size of the burst sample; the dependence on the sky coverage
is almost exclusively through the livetime. Thus care must be taken in
comparing different size samples. 

Some studies start by modeling the repetitions of burst sources, regardless of
whether the bursts are observed.  In their likelihood analyses of models with
repetition Quashnock (1995a,b) and Graziani \& Lamb (1995) assumed that each
repeating source bursts the same number of times during the observation period
(the model in \S 3).  To demonstrate the effect of sky coverage, Meegan et al.
(1995a) used this model with each source bursting 10 times during the
observation period; for $f=1/3$ and $f=1.4$ of the 1B and 2B-1B catalogs this
corresponds to $r\tau=2.5$ and 3.33, respectively, when by Figure~1 $f_r$ will
be near its asymptotic value.  Hakkila et al. (1995) used the stochastic burst
model of \S 2 and a sky coverage which varies with declination to understand
the effects of location errors and sky coverage on the repeater signal. 

Other studies model the observed repetitions without regard for the intrinsic
behavior of the source population.  Some assume that repeaters constitute a
fixed fraction of a burst sample. Strohmayer et al. (1994) generated model
catalogues by giving each burst a location, with a fraction $f_r^\prime$ being
assigned the location of a previous burst.  Thus a burst early in the catalogue
is more likely to have a repetition than a burst later in the catalogue; it is
not clear whether the resulting observed pattern is physically realizable.
Meegan et al. (1995a) constrained the possible repeater signal in various
subsets of the 2B catalogue with a model where repeating sources each are
observed to burst $\nu$ times (described as an average) and provided a fraction
$f$ of the observed bursts.  From all but one subset Meegan et al. found that
$f(\nu-1)^{1.2}/\nu\le 0.2$.  Note that the size of these subsets varies from
185 to 585, and by the analysis above the observed repeater fraction $f$ is
not expected to be constant if repeaters are present; the sample size must be
considered in evaluating the intrinsic repeater fraction (eq.~6 or 17) implied
by these constraints.  In addition, some of these subsets use bursts with
positional uncertainties less than 9$\arcdeg$, as suggested by Quashnock \&
Lamb (1993). Since the statistical uncertainty in position decreases with
increasing burst intensity, the samples with uncertainties less than 9$\arcdeg$
effectively have a higher burst threshold.  Less intense repetitions were not
included in these subsets, which consequently have a different repetition rate
$r$ for any repeating sources. Brainerd et al. (1995) considered a repetition
model similar to Meegan et al. in an analysis of spatial-temporal clustering,
from which they subsequently derived constraints on the true number of
repetitions in the 2B catalog (regardless of whether the bursts are observed). 

The simple models developed in \S 2 and \S 3 have two parameters---the burst
rate $r$ and the nonrepeater to repeater fraction $R/N_S r$---and thus an
observable such as $f_r$ from a burst sample does not fully determine the model
parameters.  For example, in Figure~3 I show the possible values of $R/N_S r$
and $r \tau$ for observed values of $f_r$ using the stochastic model of \S 2. 
As can be seen, many repetitions (large $r\tau$) from a small number of
repeating sources or few repetitions from many repeaters can result in the same
observed repeater fraction $f_r$. With more observables the parameters can be
determined; for the stochastic model, $\langle n_{obs} \rangle$ determines
$r\tau$, although it may be difficult to determine $r\tau\le 1$ since $\langle
n_{obs} \rangle$ is very nearly constant at a value of 2 for $r\tau$ in this
range.  Similarly, the dependencies of the observables on $r\tau$ (or its
equivalent) are very nearly the same for the two models developed here, and
probably for most constant rate models; it may be very difficult to identify
the repetition pattern from observations, particularly if the sky coverage is
low.  Of course, the repetition pattern can be determined if specific bursts
can be attributed to the same source. 

To prove the existence of repeaters the observed repeater fraction $f_r$ must
be shown to be nonzero.  On the other hand, to constrain the allowed repeater 
population, limits on the actual repeater fraction $F_r$ must be derived from 
the data.
\section{SUMMARY}
I have considered the observational consequences of different patterns of 
burst repetition.  I conclude:

1.  There is a fundamental difference in the sensitivity to the sky coverage 
between sources which repeat in active phases separated by time scales greater
than the observation period and sources which burst at a constant average rate.

2.  For models with sources which burst at a constant average rate the 
repeater content of a burst sample depends predominantly on the number of 
bursts in the sample; the sky coverage affects the observations primarily 
through the livetime, unless the time scales of burst repetition and the
sky coverage are correlated.

3.  Consequently, similarly-sized burst samples should be compared or a 
repetition model should be used.  Such a model is necessary to constrain the
source population based on the observations.

4.  Two models are developed in detail.  In the first each source bursts
stochastically, while in the second each source bursts at a fixed rate.  

5.  A given observed repeater fraction less than one can result from a small
number of repeating sources which burst frequently, or a larger number which
burst less frequently.
\acknowledgments
I thank R.~Lingenfelter for insightful discussions, and M.~Briggs and
J.~Hakkila for comments on this paper. This work was supported by NASA contract
NAS8-36081.

\clearpage

\figcaption{Repetition quantities as a function of the mean number of observed
bursts per source ($r\tau$) for the stochastic model.  All sources have the
same probability of bursting per unit time, and no nonrepeating sources are
included (the fraction of all bursts which originate on repeaters, whether or
not the bursts are observed, is $F_r=1$).  The number of observed bursts from
sources observed to repeat (i.e., from which there are two or more observed
bursts) is $n_{obs}$ (long dashes), while $\langle N_{S,obs} \rangle/N_S$ (dots
and dashes) is the fraction of the sources from which repetitions are observed.
The fraction of the observed bursts from sources with two or more observed
bursts is $f_r$ (solid), while $f_r^\prime$ (short dashes) is the fraction of
the bursts which are repetitions of an earlier burst.  Note that the first of a
series of two or more observed events is included in $f_r$ but not in
$f_r^\prime$.} 

\figcaption{Repetition quantities as a function of the number of observed
bursts per source ($f_s n_B$) for constant rate models. In any observation
period all sources are assumed to burst the same number of times $n_B$.  Points
(the number of bursts per source is an integer) for $f_s=1/4$ (asterisks) and
$f_s=1/3$ (pluses) are shown. Labels indicate the points for three different
quantities: $n_{obs}$---the number of observed bursts from sources observed to
repeat; $\langle N_{S,obs} \rangle/N_S$---the fraction of the sources from
which repetitions are observed; and $f_r$---the fraction of the observed bursts
from sources with two or more observed bursts.} 

\figcaption{Nonrepeater to repeater fraction as a function of the observed
number of bursts per source for different values of the repeater fraction 
$f_r$ in the stochastic model (\S 2).  Each curve is labeled by the value of
$f_r$.}


\clearpage

\begin{thebibliography}{}
%
\bibitem[Brainerd et al. (1995)]{brai95}Brainerd,~J.~J., Meegan,~C.~A.,
Briggs,~M.~S., Pendleton,~G.~N., \& Brock,~M.~N. 1995, \apjl, 441, L39
%
\bibitem[Fishman et al. (1994)]{fish94}Fishman,~G.~J., et al. 1994, \apjs,
92, 229
%
\bibitem[Graziani \& Lamb (1995)]{graz95a}Graziani,~C., \& Lamb,~D.~Q. 1995b, 
in High Velocity
Neutron Stars, ed. R.~Rothschild \& R.~Lingenfelter (New York: AIP), in press
%
\bibitem[Hakkila et al. (1995)]{hakk95}Hakkila, J., Vo,~V., Meegan,~C.,
Horack,~J., Fishman,~G., Hartmann,~D., Pendleton,~G., Briggs,~M., \&
Paciesas,~W. 1995, in Towards the Source of Gamma-Ray Bursts, 29th ESLAB
Symposium, ed. K.~Bennett \& C.~Winkler, Astr. \& Space Sci., 231, 23
%
\bibitem[Quashnock (1995a)]{quas95a}Quashnock, ~J.~M. 1995a, in High Velocity
Neutron Stars, ed. R.~Rothschild \& R.~Lingenfelter (New York: AIP), in press
%
\bibitem[Quashnock (1995b)]{quas95b}--------- 1995b, in 
Towards the Source of Gamma-Ray Bursts, 29th ESLAB Symposium, ed. K.~Bennett 
\& C.~Winkler, Astr. \& Space Sci., 231, 23
%
\bibitem[Quashnock \& Lamb (1993)]{quas93}Quashnock, ~J.~M., \& Lamb,~D.~Q. 
1993, \mnras, 265, L59
%
\bibitem[Meegan et al. (1994)]{meeg94}Meegan,~C.~A., et al. 1994, ``BATSE 2B
Gamma-Ray Burst Catalog'' available from {\tt grossc.gsfc.nasa.gov}, username 
gronews 
%
\bibitem[Meegan et al. (1995a)]{meeg95a}Meegan,~C.~A., Hartmann,~D.~H., 
Brainerd,~J.~J., Briggs,~M.~S., Paciesas,~W.~S., Pendleton,~G.~N., 
Kouveliotou,~C., Fishman,~G., Blumenthal,~G. \& Brock,~M.~N. 1995a, \apjl, 
446, L15
%
\bibitem[Meegan et al. (1995b)]{meeg95b}Meegan,~C.~A., et al. 1995b, ``BATSE 3B
Gamma-Ray Burst Catalog'' available from {\tt grossc.gsfc.nasa.gov}, username
gronews; paper in preparation 
%
\bibitem[Petrosian \& Efron (1995)]{petr95}Petrosian,~V., \& Efron,~B. 1995, 
\apjl, 441, L37
%
\bibitem[Strohmayer, Fenimore \& Miralles 1994]{stro94}Strohmayer,~T.~E.,
Fenimore,~E.~E., \& Miralles,~J.~E. 1994, \apj, 432, 665
%
\bibitem[Wang \& Lingenfelter (1993)]{wang93}Wang,~V.~C., \& 
Lingenfelter,~R.~E. 1995, \apj, 416, L13
%
\bibitem[Wang \& Lingenfelter (1995)]{wang95}--------- 1995, \apj, 441, 747
%
\end{thebibliography}
\end{document}